\documentclass[prl,showpacs,preprintnumbers,amsmath,amssymb,tightenlines,superscriptaddress,twocolumn]{revtex4}

\usepackage{graphicx}
\usepackage{dcolumn}
\usepackage{bm}
\usepackage{epsfig}
\usepackage{stmaryrd}
\usepackage{color}

\newcommand{\bra}[1]{\langle\,{#1}\, |}
\newcommand{\ket}[1]{|\,{#1}\,\rangle}

\newcommand{\ssection}[1]{{\noi  \it #1:}}

\newcommand{\sub}[2]{{#1}_{\mbox{\!\! \scriptsize #2}}}
\newcommand{\bv}[1]{\mbox{\boldmath$ #1 $}}

\def\noi{\noindent}
\def\beq{\begin{equation}}
\def\eeq{\end{equation}}

\def\CR{\nonumber\\[0.15cm]}
\newcommand{\bref}[1]{(\ref{#1})}

\newcommand{\fref}[1]{Fig.~\ref{#1}}
\newcommand{\frefp}[2]{Fig.~\ref{#1}(#2)}
\newcommand{\eref}[1]{Eq.~(\ref{#1})}

\begin{document}

\title{Non-Hermitian matter wave mixing in Bose Einstein condensates: dissipation induced amplification}
\author{S.~W\"uster}
\affiliation{Department of Physics, Bilkent University, Ankara 06800, Turkey}
\email{sebastian.wuster@bilkent.edu.tr}
\author{R.~{El-Ganainy}}
\affiliation{Department of Physics, Michigan Technological University, Houghton, Michigan 49931, USA}
\affiliation{Henes Center for Quantum Phenomena, Michigan Technological University, Houghton, Michigan 49931, USA}
\email{ganainy@mtu.edu}
\begin{abstract}
We investigate the nonlinear scattering dynamics in interacting atomic Bose-Einstein condensates under non-Hermitian dissipative conditions. We show that by carefully engineering a momentum-dependent atomic loss profile one can achieve matter-wave amplification through four wave mixing in a one-dimensional quasi free-space setup - a process that is forbidden in the counterpart Hermitian systems due to energy mismatch.  Additionally, we show that similar effects lead to rich nonlinear dynamics in higher dimensions. 
Finally, we propose a physical realization for selectively tailoring the momentum-dependent atomic dissipation. Our strategy is based on a two step process: (i) exciting atoms to narrow Rydberg- or metastable excited states, and (ii) introducing loss through recoil; all while leaving the bulk condensate intact due to protection by quantum interference.
\end{abstract}


\maketitle

\begin{figure}[htb]
\centering
\epsfig{file={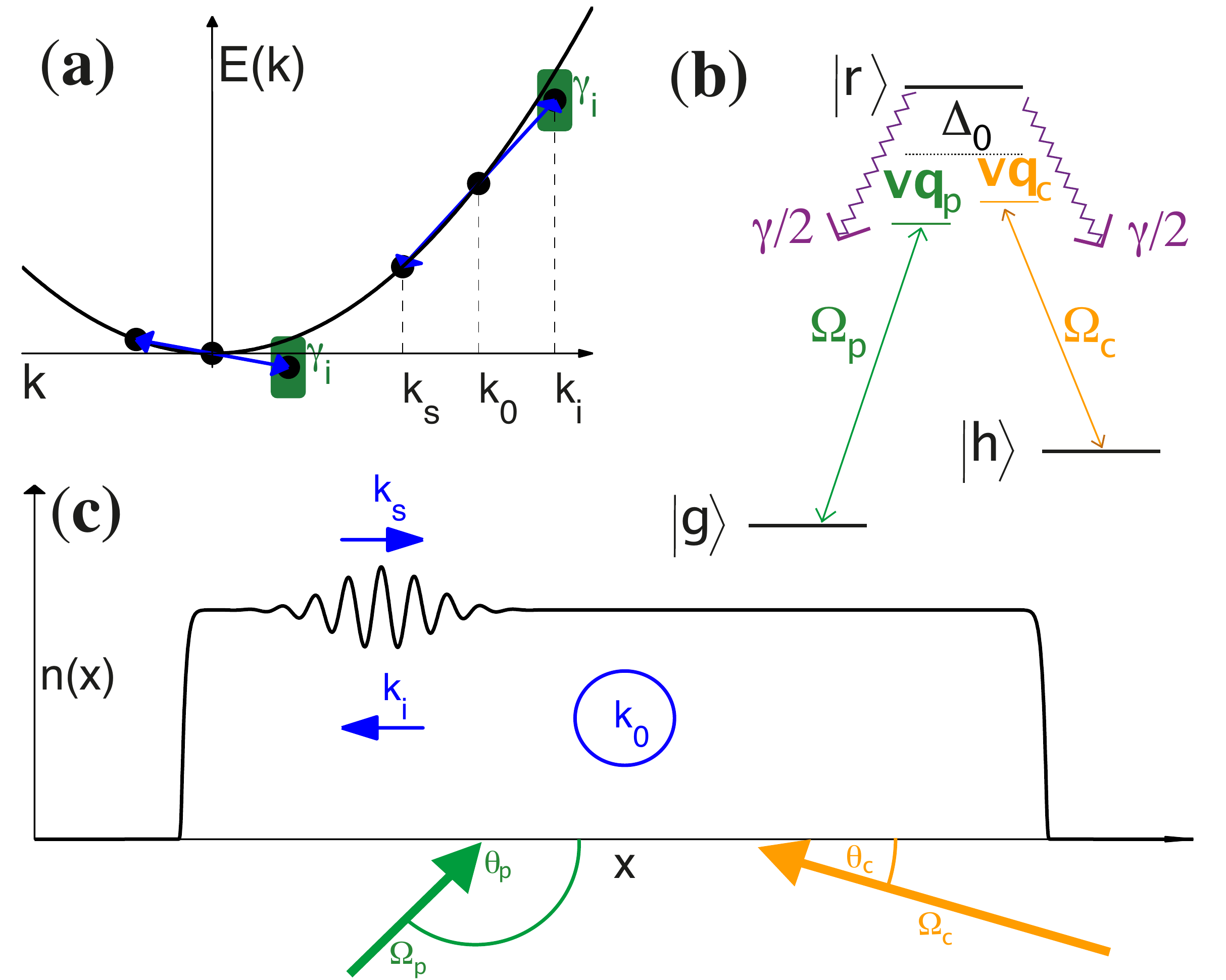},width=\columnwidth} 
\caption{(a) The non-linear dispersion relation $E(k)=\hbar^2 k^2/(2m)$ of matter waves usually prohibits one-dimensional scattering processes of the kind $2 k_0 \rightarrow k_s+k_i$ as shown. This can be remedied by engineering a loss induced width $\gamma_i$ of the final scattering state (green). (b) Internal $\Lambda$-type level scheme of the atoms used for engineering the loss, as discussed in the text. (c) Matter-wave packet (signal) of centre wavenumber $k_s$ within a 1D BEC (pump) at rest ($k_0=0$). Laser induced loss only for wave numbers ($\sim$ momenta) near $k_i=-k_s$ (idler) cause the amplification of the signal through slight depletion of the pump. $\theta_{p,c}$ are the angles of laser beams in (b) with the 1D axis of the condensate. 
\label{system_sketch}}
\end{figure}

\ssection{Introduction}
%
Wave mixing is a fundamental process associated with nonlinear interactions that involve several wave components. The most common wave mixing processes involve three and four interacting waves. In optics, both of these processes occur \cite{book:boyd_nonlinopt}, and are utilized in many applications ranging from second harmonic generation and parametric amplification to the generation of entangled photons and squeezed light. In atomic Bose-Einstein condensates (BEC), nonlinear interactions naturally lead to four wave mixing (4WM)  \cite{Inouye_phasecoh_amplific,Krachmalnicoff_4wmquantum_prl,deng_4wm_matterwaves_nature,Trippenbach_4wmtheory_pra} - a feature that is employed to generate entangled atomic beams \cite{twinbeams_buecker_nature,buecker_dynampl_pra,gross_homodyne_twinbeam_nature}.   An efficient wave mixing- or scattering process must satisfy energy and momentum conservation. This poses severe limitations for one-dimensional systems, which necessitate dispersion engineering in optics and prohibit wave mixing in homogeneous 1D BEC setups  \cite{molmer_phasematch_periodic_NJP} as illustrated in  \frefp{system_sketch}{a}.

Inspired by recent activities on nonlinear PT symmetric photonic structures \cite{ramy:coupledPT_OL,ruter:observPToptics:natphys,Peng:PTwhispgll:natphys,ramy:exceppts:pra,ramy:supersymmarray:pra,Jin:PTsymmphonlas:prl,schoenleber:optomech:nonhermi,wimmer:PTsol,Ge:modalintPT,Teimourpour:nonhermieng:2Darray,lossyopa_ElGanainy,Wasak_4wm_coupler_OL,Antonosyan_PTamplif_OL} particularly on non-Hermitian optical parametric amplification \cite{lossyopa_ElGanainy}, we show here that non-Hermitian engineering in BEC can alleviate some of these limitations and enable a host of intriguing effects. In particular, we show that in a 4WM scattering process with degenerate input states and two distinct output modes, the introduction of a selective atomic \emph{loss in just one} of the output modes can lead to the amplification of the second output state. The physical mechanism underlying this counter-intuitive effect can be understood by recalling the Heisenberg uncertainty principle: introducing loss to one component leads to a finite energy uncertainty associated with that state and thus allows for additional flexibility in the fulfilment of energy and momentum conservation (see Fig.1). Interestingly, this strategy opens up 4WM channels in 1D BEC setups, which otherwise exist only through dispersion engineering via optical lattices \cite{molmer_phasematch_periodic_NJP,campbell_paramp_lattice_prl,Hilligsoe_phasematch_PRA} or by invoking internal degrees of freedom to fulfill energy conservation \cite{klempt_paramp_spinor_prl}. 

As we will show later, the required momentum-dependent atom loss can be engineered by coupling the condensate atoms to rapidly decaying electronic states, allowing these atoms to be ejected from the trap via photon recoil. Momentum selectivity enters via the Doppler effect in a similar fashion as in laser cooling. To implement these features, we adapt a cooling scheme \cite{morigi:cooling:prl,morigi:cooling:pra,roos:EITcooling:expt:prl} based on electromagnetically induced transparency (EIT) \cite{fleischhauer:review}, by exchanging low-lying excited states with energetically narrow highly excited Rydberg states \cite{Mohapatra:coherent_ryd_det,Mohapatra:giantelectroopt} or metastable states \cite{mauger:strontium_rydberg_eit,Hodgman:metastable_lifetime_prl}. Their small natural line-widths are required to resolve the small atomic momenta and hence Doppler-shifts in the BEC, while EIT quantum interference protects the bulk condensate at zero velocity from loss. 

As our main result, we demonstrate that certain loss profiles $\gamma(\bv{k})$ can lead to the amplification of a phonon wavepacket in homogeneous 1D condensates through wave mixing. Afterwards, we investigate the collisions of three separate condensates \cite{deng_4wm_matterwaves_nature}  under non-Hermitian conditions and demonstrate novel features associated with loss-induced scattering channels. 
These additional channels may enhance the utility of BEC for atom interferometry \cite{haine_atominf_4wm}, atom-lasers \cite{dall:paired_beam,wasak_atomlaser4wm_pra} or entanglement generation \cite{perrin_atom4wm_njp,ferris_entanglem_det_lattice,Perrin_atompairexpt_4wm_prl,Jaskula_subpoiss_expt_prl}. Finally, we provide a specific example of how the required loss profiles can be experimentally engineered.

\ssection{Non-Hermitian four-wave mixing}
%
We start by assuming a 1D BEC made of atoms with mass $m$. Within the mean field Gross-Pitaevskii-equation (GPE), the system can be described in the momentum representation:
\begin{align}
&i\hbar \frac{\partial \phi(k)}{\partial t}=\frac{\hbar^2 k^2}{2 m}\phi(k) + U \int_{-\infty}^{\infty} dk_1\int_{-\infty}^{\infty} dk_2\int_{-\infty}^{\infty} dk_3
\CR
 &\times \delta(k_2 + k_3 -k_1 -k) \phi^*(k_1)\phi(k_2)\phi(k_3)  - i \hbar \gamma(k) \phi(k), 
\label{GPEmom}
\end{align}
where $\delta(k)$ is a Dirac delta function expressing the conservation of momentum during atomic (s-wave) scattering processes with interaction constant $U$. We also include momentum dependent loss with rate $\gamma(k)$. The components of the condensate momentum space wave function $\phi(k)$ at different momenta $k_i$ can couple through atomic scattering processes, possibly giving rise to wave mixing.

To first gain insight into the effect of loss on wave mixing processes, we consider a simplification of the non-linear many-mode problem \bref{GPEmom}, where only three discrete momenta $k_0$, $k_s$, $k_i$ are involved in a scattering process as shown in \fref{system_sketch}. These are referred to as pump-, signal- and idler modes respectively, in analogy to the optical scenario \cite{lossyopa_ElGanainy}.
Furthermore, we assume that only atoms in mode $k_i$ experience losses with rate $\gamma = \gamma(k_i)$.
Nextly, we express the momentum amplitudes in an interaction picture, $\Phi_{n}=\phi(k_n)\sqrt{\Delta k}\exp{[i (E_n + \epsilon_n) t/\hbar]}$, with $E_n=\hbar^2 k_n^2/(2m)$ and $\epsilon_n=\{0,\hbar\kappa/2,-\hbar\kappa/2 \}$ for $n\in\{0,s,i\}$. Here $\Delta k$ is a wave-number scaling factor, chosen as the width of the discrete momentum mode at $k_n$, and $\kappa =({\Delta}E -U \rho)/\hbar$, with energy mismatch ${\Delta}E=E_s+E_i -2E_0 $, for a homogeneous condensate with density $\rho$. Considering an isolated wave mixing process involving only $k_{0,s,i}$ rather than more general complex many-mode interactions as in \bref{GPEmom} is justified under the condition $\Delta E \gg U\rho$ \cite{footnote:fewvsmanymodes}.

It is now straightforward to show \cite{sup:inf}, that as long as the bulk BEC at $k_0$ remains largely unaffected (undepleted pump approximation), we have
\begin{align}
  i\hbar \frac{\partial \Phi_s}{\partial t}&= \left( 2 U \rho + \frac{\hbar\kappa}{2}\right)\Phi_s + U \rho\: \Phi_i^*,
  \CR
  i\hbar \frac{\partial \Phi_i}{\partial t}&=  -\left( 2 U \rho + \frac{\hbar\kappa}{2}\right)\Phi_i - U \rho\: \Phi_s^*   -i\hbar \gamma \Phi_i.
\label{GPE_disc_twomodes_redef}
\end{align}
Eq.~\bref{GPE_disc_twomodes_redef} admits solutions of the form $\bv{\Phi}(t)=\bv{\Phi}^{(+)} \exp[-i \lambda^{(+)} t] +  \bv{\Phi}^{(-)} \exp[- i \lambda^{(-)} t]$, where $\bv{\Phi}(t)  = [ \Phi_s , \: \Phi_i]^T$ and the eigenvalues
\begin{align}
\lambda^{(\pm)} =-i \frac{\gamma}{2} \pm \frac{1}{2} \bigg[-\gamma^2 + 4( 2 U \rho/\hbar + \kappa/2)^2 
\CR
- 4( U \rho/\hbar)^2 + 4 i\gamma( 2 U \rho/\hbar + \kappa/2) \bigg]^{1/2}.
\label{evals}
\end{align}
In order for amplification to take place, at least one of the above eigenvalues must satisfy Im$[\lambda^{(\pm)}]>0$.
Under the condition $\Delta E \gg U\rho$ discussed in \cite{sup:inf}, we find the imaginary part of the amplifying eigenvalue:
\begin{align}
\mbox{Im}[ \lambda^{(+)}]=\frac{(U\rho/\hbar)^2 \gamma}{\gamma^2 + (\Delta E/\hbar)^2}. 
\label{growthrate}
\end{align}
From \eref{growthrate} one can make the following important observations: (i) $\mbox{Im}[ \lambda^{(+)}]$ is maximal when $\gamma\approx \Delta E/\hbar$. Intuitively, the loss then broadens the energetic width of the idler state just enough to satisfy energy conservation, as shown in \fref{system_sketch}.   (ii) The condition $\Delta E \gg U\rho $ implies $ \mbox{Im}[ \lambda^{(+)}] \ll U\rho/\hbar $. Since $U\rho/\hbar $ sets the time-scale for non-linear BEC mean-field dynamics, the amplification will take place at a slower pace than the interaction dynamics. We refer to \cite{sup:inf} for a numerical validation of the discussion so far.

\begin{figure}[htb]
\centering
\epsfig{file={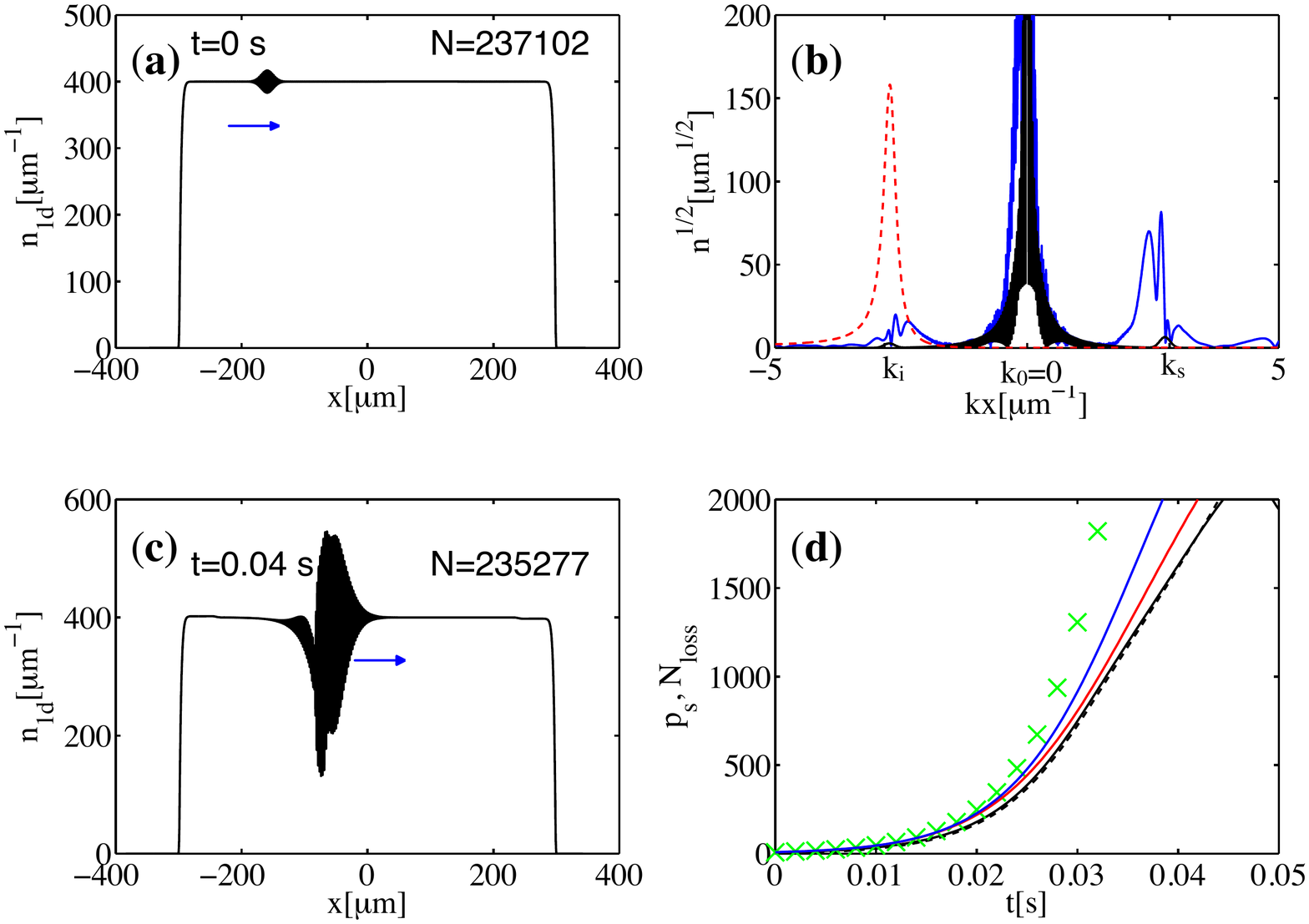},width=\columnwidth} 
\caption{1D matter wave parametric amplification, with parameters \cite{footnote:params_fig2}. (a) Initially a small wave-packet on a large top-hat shaped condensate wavepacket is travelling near wavenumber $k_s$ to the right. (b) Momentum spectra of condensate at $t=0$ (black) and $t=0.04$s (blue) as well as loss-spectrum $\gamma(k)$ in arbitrary units (red dashed). See \cite{sup:inf} for a derivation of the line-shape. (c) After $0.04$s action of loss on the wavenumber $k_i=-k_s$, the signal has been significantly amplified, with intensity drawn from the background (pump) condensate near $k_0\approx0$.
(d) The signal strength $p_s=\int_{k_s- K}^{k_s+ K} |\phi(k)|^2 dk $ (solid black) increases in exact correspondence to the number of lost atoms $\sub{N}{loss}=N(0)-N(t)$ (dashed black). Suitably chosen \cite{footnote:match_homog} simulations for a homogeneous condensate give similar results: Eq.~6 of the supplement (solid blue) and \eref{GPEmom} (solid red). Finally we also show the analytical solution of \eref{GPE_disc_twomodes_redef} using \bref{evals} ($\times$ green). See also supplementary movie.
\label{signal_amplification}}
\end{figure}

\ssection{Matter-wave signal amplification}
%
We proceed to demonstrate the non-linear amplification of a matter wave, enabled by dissipation, in a finite size multi-mode system. We consider an approximately homogeneous, one-dimensional, $^{87}$Rb BEC in a box trap, such as e.g.~in experiment \cite{Meyrath_boxBEC_PhysRevA}, with length $L=640$ $\mu$m and transverse trapping frequency $\omega_{\perp}=(2\pi)100$ Hz.
This simplifies the numerical solution of \eref{GPEmom}, but more importantly represents a regime where, in the absence of loss, matter-wave mixing processes are suppressed in one-dimension. Starting from the ground-state of the condensate in the box trap, we imprint a small "signal" matter wave packet, $\phi(x)\rightarrow \phi(x)+A_s \exp{[-(x-x_0)^2/(2\sigma)+ i k_s x]}$, with amplitude $A_s$ and width $\sigma$ centered around $k_s=2.7$ $\mu$m$^{-1}$ onto the BEC, as shown in \frefp{signal_amplification}{a}. The thick part of the line are unresolved fast oscillations with wavelength $\lambda_s=2\pi/k_s$.

Panel (b) depicts the initial wave packet of \frefp{signal_amplification}{a} in Fourier space, where we distinguish the sinc shaped bulk condensate peak ("pump", $k_0$) and the small signal near $k_s$.
We now assume the loss is switched on at time $t=0$, with profile $\gamma(\bv{k})$ shown as red dashed curve in panel (b). Note, that the spectral distribution of the loss is centered around $k_i=-k_s$. We will discuss later how to obtain such a profile. Action of this loss results in a significant amplification of the matter wave signal as shown in panel (c), together with visible bulk condensate depletion behind the passing signal wave-packet. In momentum space this manifests as rapid growth of population in momentum modes around $k_s$ as shown in panel (d). At later times, the amplification triggers complicated non-linear multi-mode dynamics, as we can see in the supplementary movie \cite{sup:inf}. We note that in the absence of dissipation, none of these effects would take place and instead the initial wave packet of panel (a) would bounce periodically off the box edges without any change in its amplitude. 

For comparison, we added to panel (d) the signal mode growth rate predicted by \eref{evals} for a homogeneous condensate closely matching the present scenario in the three-mode approximation ($\times$ green), which agrees well with full numerical solutions. Eqns.~\bref{evals} and \bref{growthrate} can thus provide useful guidance towards the parameters supporting non-Hermitian signal amplification. 

We emphasize that while no significant dynamics takes place without the loss, for just a small fraction of lost atoms ($\sim1\%$), a dramatic change in matter wave dynamics is visible at $t=0.04$s and even more so at later times. 

\ssection{Non-degenerate 2D case}
%
The scenario discussed above represents degenerate four-wave mixing, where two of the initial momenta of a scattering process co-incide ($k_0$).  A more general four-wave mixing process involves four different momentum components, and has been exploited for condensate collisions \cite{Thomas:condcollis,buggle:condcollis}, which lay the basis for studying EPR correlations with massive particles \cite{Kofler_EPRcollid_PRA,perrin_atom4wm_njp,oegren:atomcorrel_collid_pra}. In all these processes, conservation of energy and momentum plays a crucial role. A striking demonstration of this is the Bose stimulated creation of a new momentum component after the collision of three different condensates having distinct initial momenta~\cite{deng_4wm_matterwaves_nature}. 

\begin{figure}[htb]
\centering
\epsfig{file={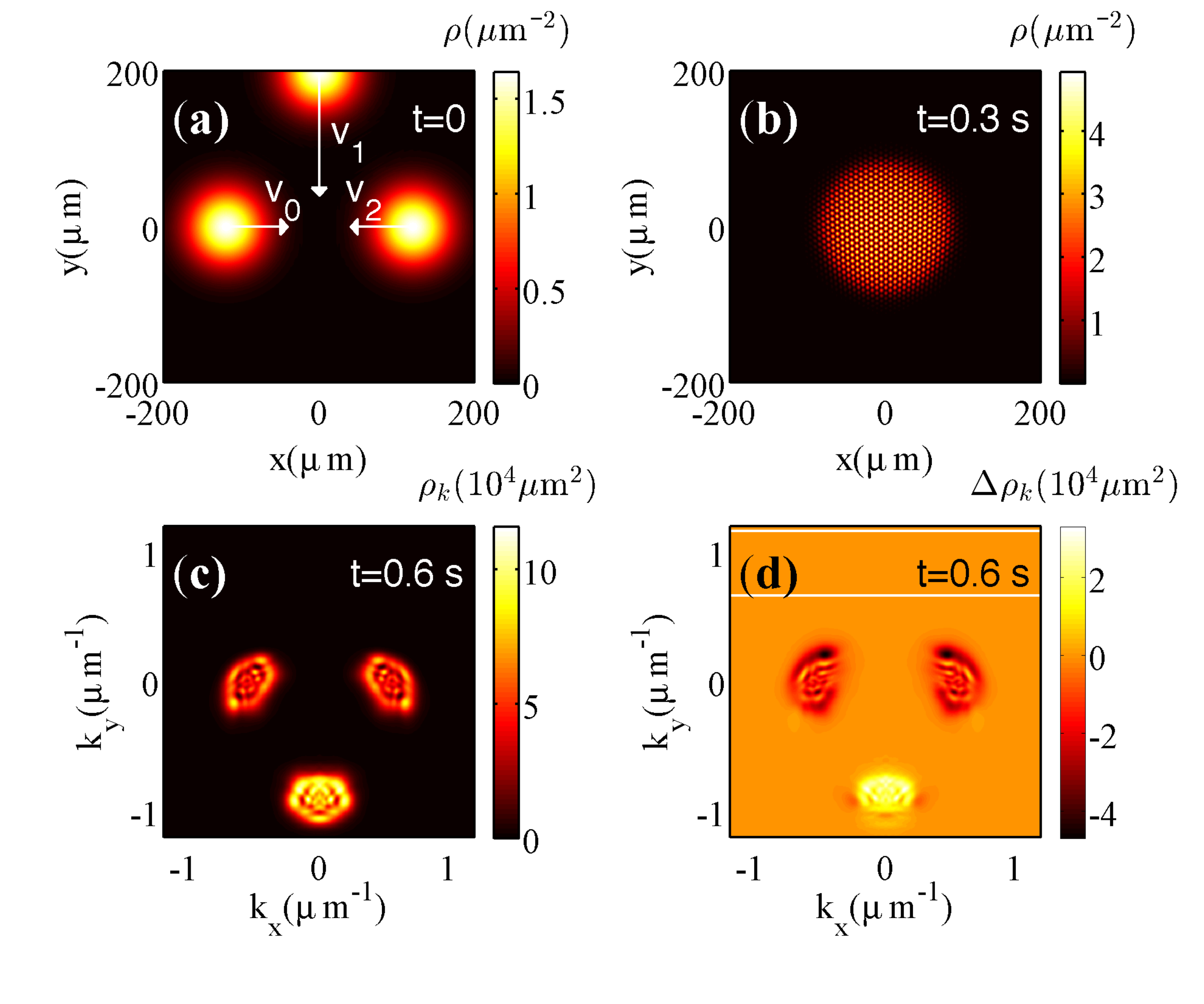},width=\columnwidth} 
\caption{Non-Hermitian effects in 2D wave mixing. (a) 2D condensate density $\rho=|\phi(x)|^2$ prior to mixing, all wave packets move towards the origin and reach it at the same time. (b) Density during collision, showing typical interferences.
(c) Resulting momentum space density $\rho_k=|\phi(k)|^2$ after the collision in the presence of loss. (d) Images in (c) after subtraction of density for case without loss. The white horizontal lines indicate the FWHM of a loss-peak shaped as that in \frefp{signal_amplification}{c} in the $y$-direction and independent of $x$, using parameters \cite{footnote:params_fig3}. See also the supplementary movie.
\label{wavemix2d}}
\end{figure}
In order to illustrate how engineering the matter dissipation can seemingly relax energy-momentum constraints, we consider a 2D scattering process. \frefp{wavemix2d}{a} depicts three separate condensate clouds, with $N_0=12000$ atoms each, in a pancake trap with $\omega_\perp=(2\pi)200$ Hz, at the locations shown on the figure and with the initial velocities $|v_{0,2}|=0.4$ mm/s, $|v_1|=0.66$ mm/s as indicated. Trapping in the $x$, $y$ directions is neglected. Since e.g.~the momentum allowed scattering process $p_0+p_2 \rightarrow p_1 +  p_3$ with $p_3=-p_1$ violates energy conservation, the clouds pass each other in the absence of loss, with just diffusive- and interaction induced broadening. However, adding momentum dependent loss with a peak within the white stripes of panel (c) (i.e.~around $p_3$), introduces an energy width alleviating these constraints as discussed earlier. Consequently our simulations show a $17\%$ increase of the signal around $p_1$ after the condensate collision, originating from these stimulated scattering events. These results should be experimentally accessible.

\ssection{Spectrally selective loss}
%
It was crucial for the development so far that the loss term $\gamma(k)$ affects only atoms in the idler mode $k_i$ and not $k_{0/s}$. Here we discuss one possible method to reach that goal, exploiting Doppler-shift techniques borrowed from laser-cooling, which also hinges on velocity selective manipulations of atoms. We describe the scheme here briefly, with more details in \cite{sup:inf}. Consider BEC atoms in their electronic ground-state $\ket{g}$, which are off-resonantly driven by a probe laser with Rabi-frequency $\Omega_p$ and detuning $\Delta$ into an excited state $\ket{r}$ that decays with rate $\Gamma$, as shown in \frefp{system_sketch}{b}. In contrast to laser-cooling, we assume that the photon recoil energy $E_r$ of decayed atoms is larger than the trap depth - a condition that can be fulfilled experimentally.  Consequently, an atom emitting a photon will be ejected out of the trap and is thus considered lost. Since the motion of condensate atoms yields a Doppler-shifted laser frequency, and ignoring state $\ket{h}$ for the moment, we obtain a loss profile $\gamma(\bv{k}) = \Gamma \: \rho_{rr}(\Delta - \bv{k}\cdot \bv{q}_p)$, where $\bv{q}_p$ is the wave-vector of the probe laser and $\rho_{rr}$ the steady state population in level $\ket{r}$. The profile $\rho_{rr}( \bv{k})$ is usually the well known Lorentzian spectral line-shape of state $\ket{r}$. 

For our scheme to function as intended, the width $\Delta k$ of $\rho_{rr}$ must be tailored such that the atomic loss is significant only for the idler matter waves $k_i$, but negligible on the bulk background condensate $k_0$. This is challenging since the Doppler shifts caused by velocities such as used in \fref{signal_amplification} are as small as $\delta \omega = v_s \cdot q_p \approx 40$ kHz. We are thus led to the use of "long-lived" excited states for $\ket{r}$, such as Rydberg states \cite{Mohapatra:coherent_ryd_det,Mohapatra:giantelectroopt} or metastable spin triplets in two-electron atoms \cite{mauger:strontium_rydberg_eit,Hodgman:metastable_lifetime_prl} which have line-widths of this magnitude. The ensuing combination of Rydberg- and BEC physics hold promise as an exciting emerging discipline \cite{balewski:elecBEC,karpiuk:imagingelectrons,nils:supersolids,moebius:cat,mukherjee:phaseimp,leonhardt:unconstrained,wang:rydelecBEC}.

However even for these states, the strong tails of the Lorentzian can cause a significant loss of bulk condensate atoms near $k_0\approx 0$. In order to overcome this obstacle, the excitation scheme can be modified to include coupling $\ket{r}$ to another hyperfine ground state $\ket{h}$ with laser parameters ($\Omega_c$, $\Delta_c$). The resulting 
$\Lambda$-type level scheme, shown in \frefp{system_sketch}{b}, enables a complete suppression of non-Doppler shifted excitation (loss) at $k\approx0$ via quantum interference effects known as electromagnetically induced transparency (EIT). While matter loss at the signal momenta $k_s$ cannot be fully suppressed in this scheme, it can be made sufficiently small.

In addition to the mechanism of atom loss described above, the laser beams will also cause a dispersive energy shift $\delta E(k)$, as discussed in \cite{sup:inf}.
Though this contribution can be made small enough to have a minor effect on condensate dynamics, it is not entirely negligible. We thus included it in the simulations above as a modification of atomic dispersion $\sim \delta E(k) \phi(k)$ on the rhs.~of \eref{GPEmom}. While it does slightly affect BEC dynamics, the amplification phenomena discussed are entirely due to the dissipative contribution.

The numerical results presented here both utilize a loss spectrum that can be created through the presented scheme for realistic parameters.

\ssection{Conclusion and outlook}
%
We have proposed a mechanism, based on spectrally engineered matter dissipation, to control the nonlinear scattering dynamics in BEC systems. More specifically, we have demonstrated that by introducing a momentum dependent loss profile, scattering processes in certain directions can be enhanced. When applied to 1D BECs, our strategy enables efficient wave mixing in regimes that would have been inaccessible under Hermitian conditions. Similarly, we have demonstrated that spectrally engineered dissipation can be used to open new scattering channels in 2D setups. 

We emphasize that our primary results, namely efficient four-wave mixing via matter loss, are valid in general and not pertinent to the examples studied here. Also alternative practical realisations of loss profiles $\gamma(\bv{k})$ would yield the same results.

The fundamental effects presented in our work may have far reaching consequences for engineering quantum-atom-optical devices, such as interferometers or entangled atom sources. Along these lines, it would be interesting to investigate how the loss impacts the quantum correlations. Additionally, interesting new features might arise from inelastic non-linear loss processes in condensates, such as two- and three-body losses \cite{jila:nova,wuester:nova,wuester:nova2,wuester:phonons}. 

\acknowledgments
R.E. acknowledges support from the Henes Center for Quantum Phenomena at Michigan Technological University.


\section{Supplemental information}

\section{Non-Hermitian four-wave mixing}
 
This section presents a detailed derivation of the analytical treatment of wave-mixing in the presence of losses, leading to Eq.~(4) of the main article.
 
 For completeness, we begin with the position space GPE of a 1D BEC in a transversely tight trap by considering $N=\int dx |\Psi(x)|^2 $ atoms of mass $m$ in one spatial dimension and within an external trapping potential $V(x)$:
\begin{align}
i\hbar \frac{\partial \Psi(t,x)}{\partial t}&= \left[ -\frac{\hbar^2 }{2 m}\frac{\partial^2 }{ \partial x^2} + V(x) +  U |\Psi(t,x)|^2  \right]\Psi(t,x).
\label{GPEpos}
\end{align}
 Here $U$ is an effective 1D nonlinear coefficient, containing e.g. properties of tight transverse trapping. It can be calculated from the 3D interaction constant $\sub{U}{3D} =4 \pi \hbar^2 a_s/m$, where $a_s$ is the s-wave scattering length, via $U=\sub{U}{3D} /(2\pi\sigma_{\perp}^2)$, where $\sigma_{\perp}$ is the width of the harmonic-oscillator ground-state in the tightly trapped transverse direction $\sigma_{\perp} = \sqrt{\hbar/(m\omega_\perp)}$, and $\omega_\perp$ is the trapping frequency of the transverse trap. 

In order to facilitate our analysis, it is more instructive to write \eref{GPEpos} in the momentum representation after neglecting the potential term $V(x)$. We use $\Psi(t,x) = \int dk \exp{[i k x]} \bar\phi(t,k)/\sqrt{2\pi}$ to obtain:
\begin{align}
i\hbar \frac{\partial \bar\phi(t,k)}{\partial t}&=\frac{\hbar^2 k^2}{2 m}\bar\phi(t,k) + \frac{U}{2\pi} \int_{-\infty}^{\infty} dk_1\int_{-\infty}^{\infty} dk_2\int_{-\infty}^{\infty} dk_3
\CR
 &\times \delta(k_2 + k_3 -k_1 -k) \bar\phi^*(t,k_1)\bar\phi(t,k_2)\bar\phi(t,k_3), 
\label{GPEmom}
\end{align}
where $\delta(k)$ is a Dirac delta function expressing the conservation of momentum during atomic scattering processes.

In principle, equation (\ref{GPEmom}) contains complete information about the system. However, in order to gain more insight into a specific wave mixing process such as those represented in Fig.1~(a) of the main article, we proceed by isolating few specific momentum modes. We do this by adopting a discrete representation of Eq. \bref{GPEmom}. This is done by using $ \int dk \rightarrow \sum \Delta k$ and $ \bar\phi(t,k_n)=\bar\phi_{n}/\sqrt{\Delta k}$ (we drop the explicit reference to the time variable $t$) , to obtain: 
\begin{align}
i\hbar \frac{\partial \bar\phi_p}{\partial t}&=\frac{\hbar^2 k_p^2}{2 m}\bar\phi_p + U_d \sum_{n,m,l} \delta_{k_p + k_n -k_m -k_l} \bar\phi^*_n\bar\phi_m\bar\phi_l, 
\label{GPEmom_discr}
\end{align}
where $U_d=U\Delta k/(2\pi)$ and $\delta_{k_n}=\delta(k_n)\Delta k$. In our simulations $\Delta k$ represents the momentum grid-spacing. Finally, we write \bref{GPEmom_discr} in an interaction picture via the gauge transformation $\bar{\phi}_n=\phi_n  \exp{(-i E_n t /\hbar)}$, with $E_n=\hbar^2 k_n^2/(2m)$ to get:
\begin{align}
i\hbar \frac{\partial \phi_p}{\partial t}&= U_d \sum_{n,m,l}  \delta_{k_p + k_n -k_m -k_l} 
\CR
&\times \phi^*_n\phi_m\phi_l  e^{i(E_p +E_n -E_m -E_l)t/\hbar} .
\label{GPEmom_discr_IP}
\end{align}
This equation neatly encapsulates momentum and energy conservation inherent in any atomic scattering event.
\section{Phasematching through loss}

Let us now assume a momentum dependent single-body atom loss for the BEC, by adding a term
\begin{align}
i\hbar \frac{\partial \phi(k)}{\partial t}&=\cdots - i\hbar \gamma(k) \phi(k). 
\label{GPEloss}
\end{align}
to the rhs.~of \eref{GPEmom}. After discretisation this amounts to a term $i\hbar \frac{\partial \phi_p}{\partial t}=  \cdots -i\hbar \gamma_p \phi_p$ in \eref{GPEmom_discr_IP}.

We now further restrict \eref{GPEmom_discr_IP} to just three relevant modes, $k_0$, $k_s$, $k_i$, for loss acting on atoms with the idler wavenumber $k_i$ only, and obtain:
\begin{align}
i\hbar \frac{\partial \phi_0}{\partial t}&= U_d \bigg\{2  \phi^*_0 \phi_s\phi_i  e^{-i(E_i +E_s - 2 E_0)t/\hbar} \
\CR
&+\left[ |\phi_0|^2 + 2\left( |\phi_s|^2 + |\phi_i|^2  \right) \right]\phi_0
  \bigg\} ,
  \CR
  i\hbar \frac{\partial \phi_s}{\partial t}&= U_d \bigg\{ \phi^*_i \phi_0^2  e^{i(E_i +E_s - 2 E_0)t/\hbar} \
\CR
&+\left[ |\phi_s|^2 + 2\left( |\phi_0|^2 + |\phi_i|^2  \right) \right]\phi_s
  \bigg\} ,
  \CR
  i\hbar \frac{\partial \phi_i}{\partial t}&= U_d \bigg\{ \phi^*_s  \phi_0^2  e^{i(E_i +E_s - 2 E_0)t/\hbar} \
\CR
&+\left[ |\phi_i|^2 + 2\left( |\phi_0|^2 + |\phi_s|^2  \right)\right]\phi_i  \bigg\} -i\hbar \gamma \phi_i.
\label{GPE_disc_threemodes}
\end{align}
%

\subsection{Analytical solution}
If the majority of condensate atoms occupy the pump mode $k_0$, i.e. $|\phi_0| \gg |\phi_s|, |\phi_i|$ (a condition equivalent to the undepleted pump approximation in the context of non-linear optics), the evolution equation for $\phi_0$ becomes:
\begin{align}
i\hbar \frac{\partial \phi_0}{\partial t}&= U_d |\phi_0|^2 \phi_0,
\end{align}
with the solution $\phi_0(t) = \sqrt{n_0} e^{-i U_d{n_0} t /\hbar}$, where ${n_0}= |\phi_0|^2$. Here $n_0$ is the number of atoms in the pump mode, and $\phi_0$ the pump mode amplitude. Substitution into the remaining two equations yields:

 \begin{align}
  i\hbar \frac{\partial \phi_s}{\partial t}&= U_d {n_0} \left\{  \phi^*_i e^{i(E_i +E_s - 2 E_0 - 2U_d{n_0})t/\hbar} +2 \phi_s \right\} 
  \CR
  i\hbar \frac{\partial \phi_i}{\partial t}&= U_d {n_0} \left\{  \phi^*_s e^{i(E_i +E_s - 2 E_0 - 2U_d{n_0})t/\hbar} +2  \phi_i \right\}   -i\hbar \gamma \phi_i
\label{GPE_disc_twomodes}
\end{align}
By introducing the quantity $\kappa = (E_i +E_s - 2 E_0 - 2U_d{n_0})/\hbar$ and moving to the rotating frame $\phi_s=\Phi_s \exp{(i \kappa t/2)}$, $\phi_i=\Phi_i^* \exp{(-i \kappa t/2)}$, we obtain:
\begin{align}
  i\hbar \frac{\partial \Phi_s}{\partial t}&= \left( 2 U_d {n_0} + \frac{\hbar\kappa}{2}\right)\Phi_s + U {n_0} \Phi_i ,
  \CR
  i\hbar \frac{\partial \Phi_i}{\partial t}&=  -\left( 2 U_d {n_0} + \frac{\hbar\kappa}{2}\right)\Phi_i - U {n_0} \Phi_s   -i\hbar \gamma \Phi_i.
\label{GPE_disc_twomodes_redef}
\end{align}
The solution of \bref{GPE_disc_twomodes_redef} reads:
\begin{align}
\bv{\Phi}(t)&=\bv{\Phi}^{(+)} \exp[-i \lambda^{(+)} t] +  \bv{\Phi}^{(-)} \exp[- i \lambda^{(-)} t],
\label{GPE_disc_twomodes_solution}
\end{align}
with 
\begin{align}
\lambda^{(\pm)} =-i \frac{\gamma}{2} \pm \frac{1}{2} \bigg[-\gamma^2 + 4( 2 U_d {n_0}/\hbar + \kappa/2)^2 
\CR
- 4( U_d {n_0}/\hbar)^2 + 4 i\gamma( 2 U_d {n_0}/\hbar + \kappa/2) \bigg]^{1/2}.
\label{evals}
\end{align}
and $\bv{\Phi}^{(\pm)}$ the eigenvectors associated with the eigenvalues $\lambda^{(\pm)}$.
We have used a vector notation $\bv{\Phi}(t)  = [ \Phi_s , \: \Phi_i]^T$. Evidently, non-linear amplification of atomic momentum modes in \bref{GPE_disc_twomodes_solution} takes place if either one of the eigenvalues satisfies the condition Im$[\lambda^{(\pm)}]>0$. For our present reduction of the problem to just three interacting momentum-modes to be valid, we require $\Delta E \equiv  E_i +E_s - 2 E_0) \gg (U_d{n_0})$ \cite{footnote:fewvsmanymodes}, the expression for the amplification factor then takes the form:
\begin{align}
\mbox{Im}[\lambda^{(+)}] = \frac{(U_d{n_0}/\hbar)^2 \gamma}{\gamma^2 + (\Delta \bar{E}/\hbar)^2}. 
\label{growthrate}
\end{align}
It is useful to re-express $U{n_0}$ in terms of the homogeneous position-space density $\rho$ within a domain of size $L$. We then can write
$U_d{n_0}=U {\Delta}k N_0/(2\pi) = U  (2\pi/L) N_0/(2\pi) = U\rho$, which is used in Eq.~(2)-Eq.~(4) of the main article.

From expression \bref{growthrate}, we can read off the following important results: (i) $\mbox{Im}[\lambda^{(+)}] $ is maximal when $\gamma\approx \Delta E/\hbar$. In that case loss just broadens the target momentum state enough to satisfy the energy conservation relation, as shown in Fig.~1. For the parameters of \fref{comparisons} this predicts an optimum at $\gamma=1.2$ for which the effect is indeed much faster than the other cases shown.
(ii) Since we need $\Delta E \gg U_d{n_0} $, we always have  $\mbox{Im}[\lambda^{(+)}] \ll \gamma$. This forces us to interpret the condition for wave mixing $\Delta E \gg U_d{n_0}$ as generous as possible, with $\Delta E$ just a few multiples of $U_d{n_0} $, to have the fastest possible effect. (iii) Using all these arguments together we also have $\mbox{Im}[\lambda^{(+)}]\ll U{n_0} $. Since $U_d{n_0}$ sets the time-scale for the non-linear BEC dynamics, this will make the amplification effect slower than the latter. 

 For a case such as in Fig.~2 of the main article, the last conclusion is no limitation since the bulk nonlinearity in the homogeneous case does not lead to any non-trivial evolution of the wave function. In contrast, for Fig.~3 it limits the achievable signal since for faster amplification non-linearities become to large for the matter-waves to pass through each other.

\section{Verification of results}

\begin{figure}[htb]
\centering
\epsfig{file={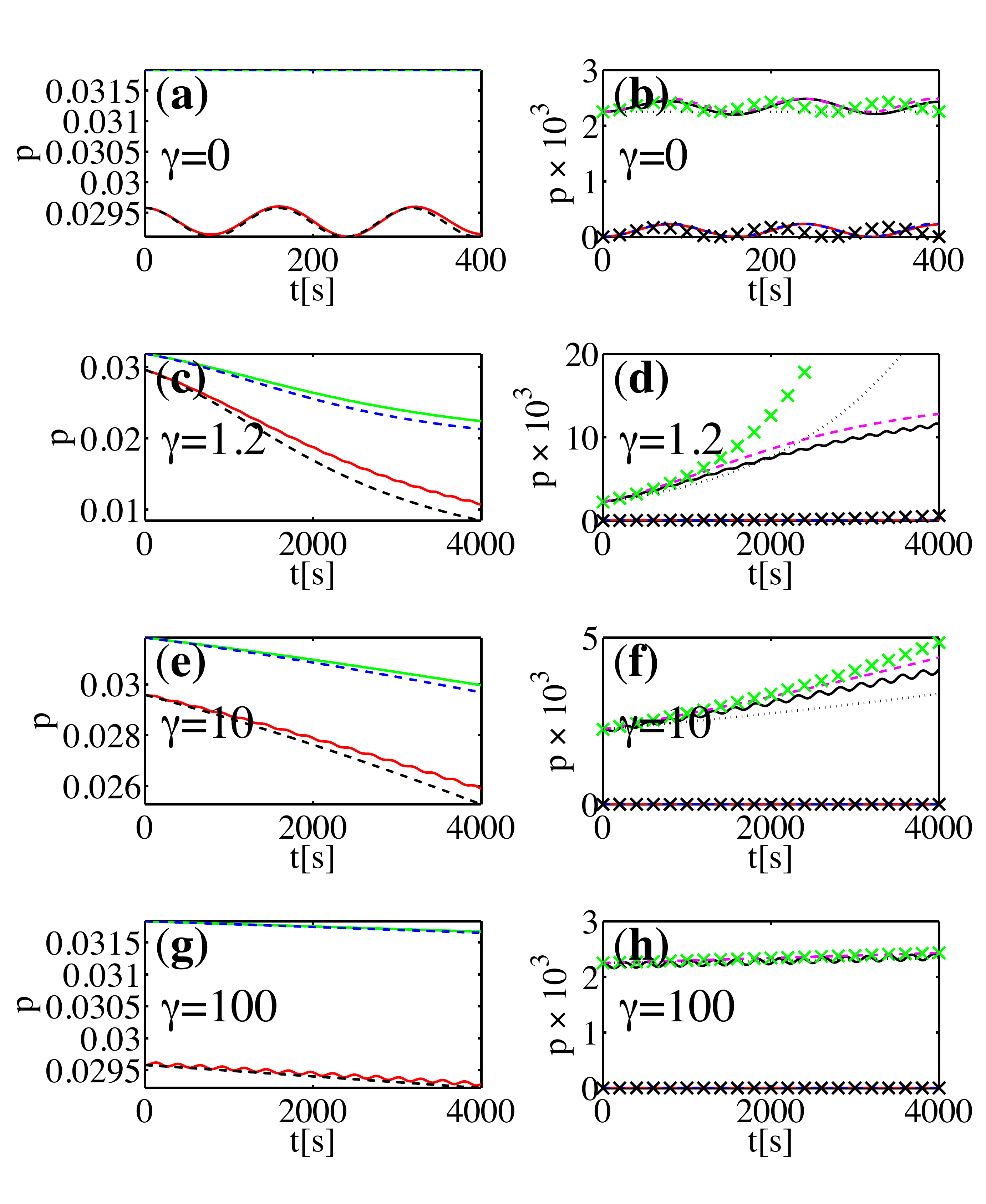},width=\columnwidth} 
\caption{Simulations of four-wave mixing in a homogenous BEC, using \eref{GPEmom} with \bref{GPEloss} (solid), compared with a few-mode approximation \eref{GPE_disc_threemodes} (dashed), and the analytical solution based on \eref{evals} ($\times$). All curves show momentum space densities $|\phi(k_i)|^2=  |\phi_{n}|^2/\Delta k$ at various wave numbers: (right panels upper curves) signal $k_s$, (right panels lower curves) idler $k_i$, (left panel lower curves) pump $k_0$, (left panel upper curves) sum of all three. From top to bottom we increase the idler damping rate $\gamma$.
\label{comparisons}}
\end{figure}

Here we present a numerical verification of the predictions made in the previous section. To do so, we consider the case of a homogeneous BEC of $N$ atoms in periodic domain of length $L$, thus with density $\rho=N/L$. The BEC is initially given a momentum $k_0$ with a small fraction $A_s$ of the BEC seeded into the signal mode (momentum $k_s$):
\begin{align}
\Psi(x,0)&=\sqrt{\frac{N}{L}} \bigg(\sqrt{1 - A_s}\exp{[i k_0 x]} + \sqrt{A_s}\exp{[i k_s x]} \bigg).
\label{GPEinistate}
\end{align}
The initial condition \bref{GPEinistate} is then propagated numerically under the loss spectrum $\gamma(k)=\gamma A \exp{(-(k-\sub{k}{loss})^2/(2 \sub{\sigma}{loss}))}$, centered around $k_i=2k_0-k_s$. Other simulation parameters were chosen similar to those listed in \cite{list_parameters}.\\
Figure \ref{comparisons} depicts the condensate momentum components at the idler and signal wavenumber for various loss rates, comparing three different models: the many mode GPE for the homogeneous case \bref{GPEmom} with \bref{GPEloss}, the corresponding three-mode model \bref{GPE_disc_threemodes}, its full analytical solution \bref{GPE_disc_twomodes_solution}. The agreement is good for this homogeneous case, as expected. We find that \bref{growthrate} can also provide useful guidance in the inhomogeneous multi-mode cases discussed in the main text.

%

\section{Engineering momentum selective loss in a BEC}

In order to experimentally observe the effects discussed so far,  the loss spectrum should act selectively only on specific momentum components of the condensate. Similar momentum dependent manipulations are essential also during BEC formation, 
where in the laser cooling stage the fastest atoms are slowed down, with velocity selectivity provided by the Doppler shift. 

There are two additional challenges here: the comparatively small velocities involved in BEC dynamics and avoiding excess loss of the majority of condensate atoms.  
We now provide some additional details on how these challenges can be met by employing a laser cooling technique that involves quantum interference, as has been proposed \cite{morigi:cooling:prl,morigi:cooling:pra} and demonstrated \cite{roos:EITcooling:expt:prl}  for the cooling of ions. 

In this scheme, condensate atoms in their ground-state $\ket{g}$ are coupled to two laser beams, in the off-resonant $\Lambda$ configuration shown in Fig.~1 (b) of the main article. The corresponding Hamiltonian for one atom in the rotating wave approximation reads
\begin{align}
\sub{\hat{H}}{EIT}= &\bigg[-\frac{\sub{\Omega}{p}}{2}\hat{\sigma}_{rg}  - \frac{\sub{\Omega}{c}}{2}\hat{\sigma}_{rh}  + \mbox{h.c.} 
\CR
&+ \Delta_p(\bv v)  \hat{\sigma}_{rr} + (\Delta_p(\bv v) - \Delta_c(\bv v)) \hat{\sigma}_{hh}    \bigg],
\label{Heit}
\end{align}
with $\hat{\sigma}_{kk'}=[\ket{k}\bra{k'}]$. The detunings $\Delta_{p,c}$ are implicitly dependent on the velocity of the atom $\bv{v}$ through the doppler shift $\Delta_{p,c}=\Delta_0 - \bv{v}\cdot \bv{q}_{p,c}$, where $\Delta_0$ is a common base detuning and $\bv{q}_{p/c}$ are the respective laser wave vectors.  For our one-dimensional simulation of Fig.~2, we in practice write this as $ \bv{v}\cdot\bv{q}_{p,c}= v |\bv{q}|a_{p,c}$, 
where $v$ is the one-dimensional velocity, $|\bv{q}|\approx |\bv{q}_p|\approx |\bv{q}_c|$ the laser wave-number and $a_{p,c}=\cos{\theta_{p,c}}$ accounts for the angle between the probe- and coupling beams and our one-dimensional condensate.

The system evolves according to the Lindblad master equation for the density matrix $\hat{\rho}$ ($h=1$)
\begin{align}
\dot{\hat{\rho}}= -i [\hat{H},\hat{\rho}] + \sum_\alpha {\cal L}_{\hat{L}_\alpha}[\hat{\rho}],
\label{mastereqn}
\end{align}
where the super-operators ${\cal L}_{\hat{L}_\alpha}[\hat{\rho}]$, with $ {\cal L}_{\hat{O}}[\hat{\rho}]=\hat{O}\hat{\rho}\hat{O}^\dagger - (\hat{O}^\dagger\hat{O}\hat{\rho}+\hat{\rho}\hat{O}^\dagger\hat{O})/2$ describe spontaneous 
decay of the excited state $\ket{r}$ to either of the ground states, via $\hat{L}_{\alpha=1}=\sqrt{\Gamma/2}\hat{\sigma}_{gr}$ and $\hat{L}_{\alpha=2}=\sqrt{\Gamma/2}\hat{\sigma}_{hr}$. 

As usual the time-scales of atom-light coupling are much faster than BEC dynamics so that we can assume the atoms to settle into a steady state $\hat{\rho}^{(\infty)}$ with $\dot{\hat{\rho}}=0$, dependent on the atomic velocity through the doppler-shift.

We finally assume that all atoms spontaneously decaying from $\ket{r}$ are lost from the trap, which yields the velocity (wave-number) dependent loss as $\gamma(k)=\Gamma \bra{r}\hat{\rho}^{(\infty)} \ket{r}$, which 
gives
\begin{align}
\gamma(k)&=8\sub{\Delta}{eff}^2 \Omega_p^2 \Omega_c^2/\bigg[  [16(\Delta_c^2 + \Omega_c^2)\sub{\Delta}{eff}^2 + 3 \Omega_c^4]\Omega_p^2
 \CR
&+(3 \Omega_c^2 -8 \Delta_c \sub{\Delta}{eff})\Omega_p^4 + \Omega_p^6 + 4 \Gamma^2 \sub{\Delta}{eff}^2 (\Omega_p^2 + \Omega_c^2)
\CR 
& + (4 \sub{\Delta}{eff} \Delta_p \Omega_c +  \Omega_c^3)^2
\bigg],
\label{gammaeff}
\end{align}
with $\sub{\Delta}{eff} = \Delta_p-\Delta_h$.

\begin{figure}[htb]
\centering
\epsfig{file={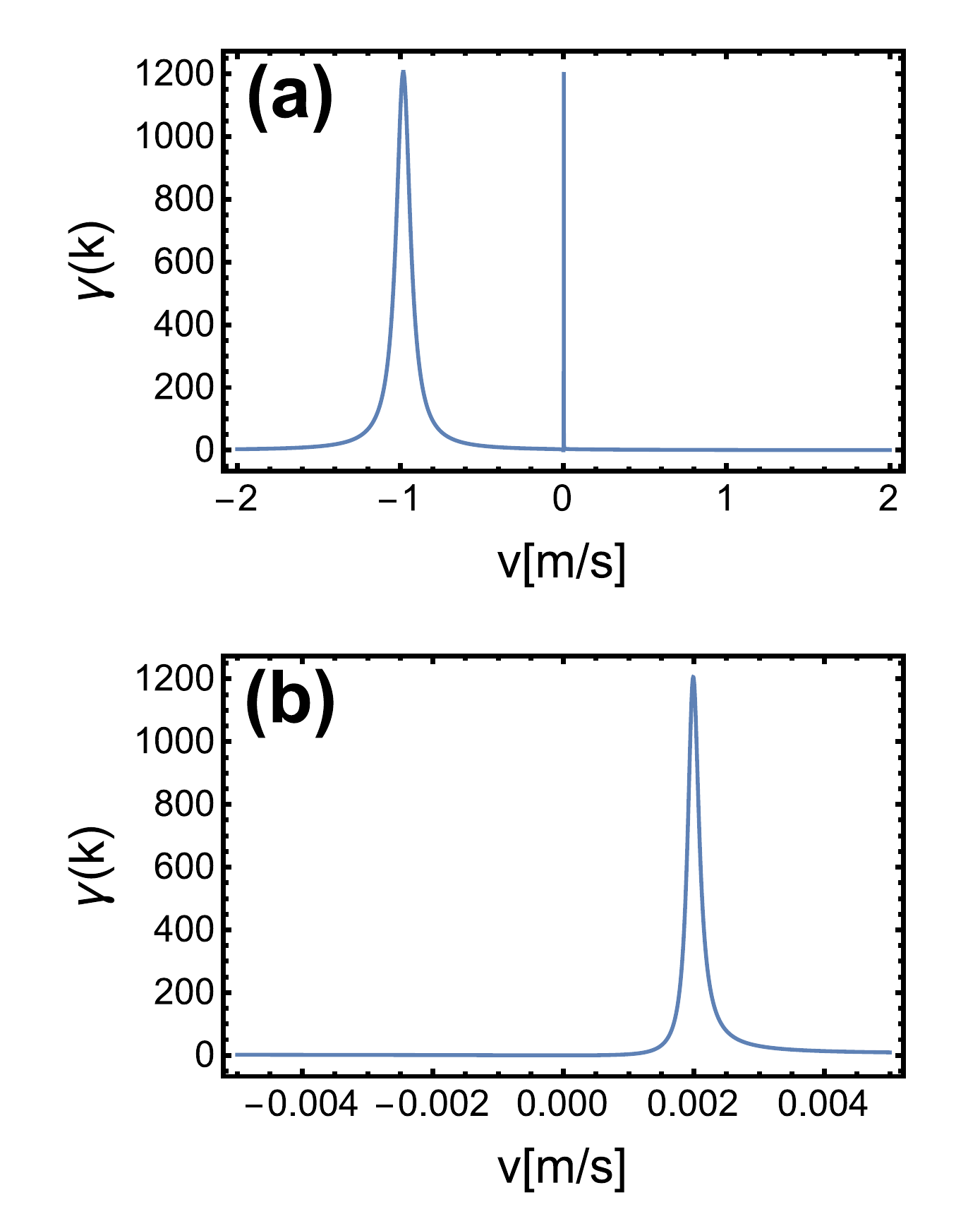},width=0.7\columnwidth} %
\caption{(a) Loss spectrum engineered through $\Lambda$ scheme as a function of atomic velocity, for parameters as in Fig.~2 of the main article.
(b) The same, but zoomed onto the narrow resonance feature near $v=v_s=0.002$ m/s. 
\label{lossspec}}
\end{figure}

We show the atomic excitation spectrum, or loss spectrum \bref{gammaeff}, as a function of velocity in \fref{lossspec}. On the wider velocity range in panel (a) we see a broad and a narrow resonance feature, corresponding to the two eigenstates of the strongly coupled $\ket{r}$, $\ket{h}$ sub-space. Our interest is in the narrow spectral feature, zoomed upon in panel (b), which realises the velocity width required for the proposal. The state is narrow due to the small probability to be in the decaying $\ket{r}$ level.

Note that in our argument we essentially rely on a hierarchy of three time-scales: $\tau \ll  \sub{T}{rec} \ll \sub{T}{BEC}$, where the excited state life time $\tau=\gamma^{-1}$ sets the scale for radiative decay establishing an atomic steady state, the recoil time $T_{rec}$ determines how fast a decaying atom is lost from the trap and $T_{BEC}$ denotes the time-scale of condensate dynamics of interest.
We estimate  $\sub{T}{rec}=a_\perp/v_{rec}$, where $a_\perp=\sqrt{\hbar/m/\omega_\perp}$ is the radial trapping width and $v_{rec} = 2\hbar |q_p|/m$ the recoil velocity. 

For the parameters of Fig.~2  in the main article we have $\tau=3.6 \times 10^{-6}$ $\mu$s, $T_{rec}=36$ $\mu$s and $\sub{T}{BEC}\sim1$ ms, fulfilling the hierarchy.

\subsection{Dispersive effects}

The optical $\Lambda$ scheme causes the ground state $\ket{g}$ to be weakly dressed with the other two electronic states. Besides the desired dissipative effects discussed above, this will cause an energy shift (light-shift) for the atoms, given by $\delta E(k) = \mbox{Tr}[\hat{\rho}^{(\infty)} \sub{\hat{H}}{EIT}]$ that also will depend on the atomic velocity. 

We find
\begin{align}
&\delta E(k)=-\delta\Omega_p^2 \bigg[
\frac{3}{2}\gamma^3 \delta^2 + \frac{1}{2}\Gamma \Omega_c^2 (4 \Delta_p \delta +  \Omega_c^2  + \Omega_c^2  )
\CR
&+\frac{\Gamma}{2} [ 4 \delta \left\{ \delta(4 \Delta_c^2 + \frac{\Gamma^2}{2}) + \Omega_c^2(3\Delta_c  - 2 \Delta_p) \right\} 
\CR
&\hspace{1cm}+ \Omega_p^2(\Omega_c^2  - 8 \Delta_c \delta) + \Omega_p^4  ]
 \bigg]
 \CR
 &\times \bigg[ 
 \frac{\Gamma \Omega_c^2}{2}[4(\Gamma^2 + 4\Delta_p^2)\delta^2 + 8 \Delta_p \delta \Omega_c^2 +  \Omega_c^4 ]
 \CR
 &+\Omega_p^2[ 2 \Gamma(\Gamma^2 + 4 \Delta_c^2)\delta^2 + 8 \Gamma\delta^2\Omega_c^2 +\frac{3}{2}\Gamma\Omega_c^4 ]
  \CR
 &+\Omega_p^4[\frac{3}{2}\Gamma\Omega_c^2 - 4 \Gamma \Delta_c \delta ] + \frac{1}{2} \Gamma \Omega_p^6
 \bigg]^{-1},
\label{Eshift}
\end{align}
with $\delta  =(\Delta_c - \Delta_p)$.

\begin{figure}[htb]
\centering
\epsfig{file={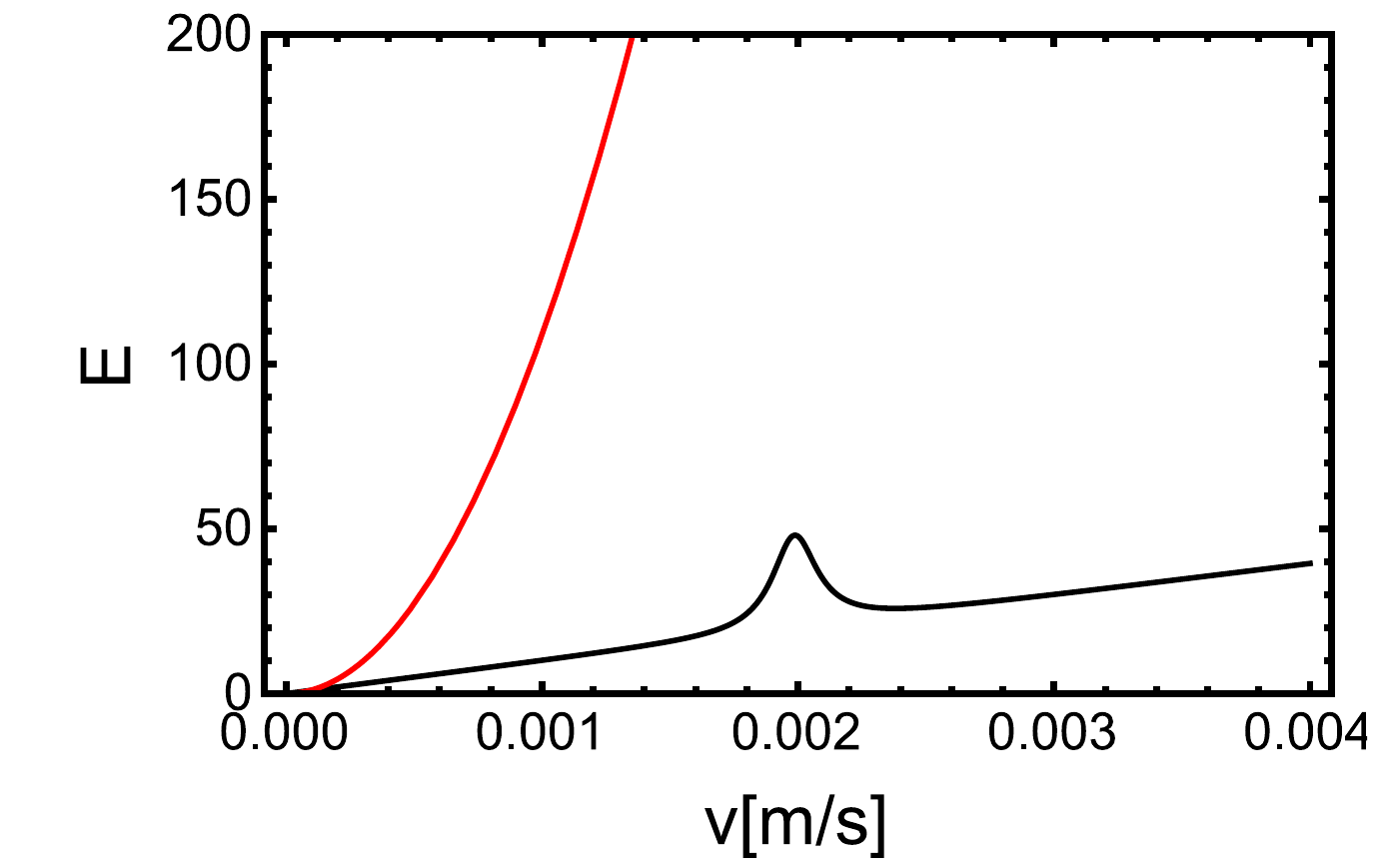},width=\columnwidth} %
\caption{(black) Light-shift from laser coupling $\delta E(k)$, \eref{Eshift}, compared to kinetic energy $\sub{E}{kin}=\frac{1}{2}m v^2$ (red). Since the former is much smaller than the latter, it does not significantly affect the condensate dynamics.
\label{dispersive_contribution}}
\end{figure}

In comparison with the free atom dispersion relation $E=\hbar^2 k^2/(2m)$ this effect remains small but not entirely negligible as seen in \fref{dispersive_contribution}.


\begin{thebibliography}{62}
\bibitem[{foo({\natexlab{a}})}]{footnote:fewvsmanymodes}
\bibinfo{note}{If $\Delta E \sim (U_d{n_0})$, the interaction energy itself can
  usually enable a much larger set of scattering channels, leading to a
  many-mode problem.}
\bibitem[{foo({\natexlab{b}})}]{footnote:params_fig2}
\bibinfo{note}{Beyond the parameters stated in the text or clear from the
  figure, we have used a phonon wave packet with $A_s=600 m^{-1/2}$, $\sigma
  =9.6\mu$m. The loss spectrum $\gamma(k)$ is based on $\Gamma=(2\pi)/\tau$
  with $\tau=3.6\mu$s (corresponding to $\ket{r}=\ket{15,p_{1/2}}$
  \cite{beterov:bbr}), $\Omega_p=(2\pi)6366$ Hz, $\Omega_c=(2\pi)405845$ Hz,
  $\Delta_0=(2\pi)3.18$ MHz at laser wavelength $\lambda=308$nm and $a_p=1$,
  $a_c=-1$.}
\bibitem[{foo({\natexlab{c}})}]{footnote:match_homog}
\bibinfo{note}{We extract the initial signal population $n_s$ from the full
  non-homogeneous simulation, and choose a spatial domain of length $L_s$ for
  the homogeneous case that matches the support of the signal wavepacket
  $L_s\approx \sigma$.}
\bibitem[{foo({\natexlab{d}})}]{footnote:params_fig3}
\bibinfo{note}{For this simulation, the loss spectrum $\gamma(k)$ was adjusted
  to $\Omega_p=(2\pi)5.6\times10^3$ Hz, $\Omega_c=(2\pi)3.45\times10^5$ Hz,
  $\Delta_0=(2\pi)7$ MHz compared to Fig.~2.}
\end{thebibliography}
\end{document}